\documentclass[a4paper,fleqn]{cas-sc}

\usepackage[authoryear,longnamesfirst]{natbib}
\usepackage{graphicx}
\usepackage{booktabs}
\usepackage{multirow}
\usepackage{longtable, array}

\usepackage{changes}

\def\tsc#1{\csdef{#1}{\textsc{\lowercase{#1}}\xspace}}
\tsc{WGM}
\tsc{QE}

\begin{document}
\let\WriteBookmarks\relax
\def\floatpagepagefraction{1}
\def\textpagefraction{.001}

\setlength{\tabcolsep}{5pt}

\shorttitle{From Passive Tool to Socio-cognitive Teammate: A Conceptual Framework for Agentic AI in Human-AI Collaborative Learning}

\shortauthors{Anonymous}  

\title [mode = title]{From Passive Tool to Socio-cognitive Teammate: A Conceptual Framework for Agentic AI in Human-AI Collaborative Learning}  



%

\author[1,2]{Lixiang Yan}[orcid=0000-0003-3818-045X]
\cormark[1]

\affiliation[1]{organization={School of Education, Tsinghua University},
            city={Beijing},
            country={China}}  

\affiliation[2]{organization={Centre for Learning Analytics, Monash University},
            city={Clayton},
            country={Australia}}          

\begin{abstract}
The role of Artificial Intelligence (AI) in education is undergoing a rapid transformation, moving beyond its historical function as an instructional tool towards a new potential as an active participant in the learning process. This shift is driven by the emergence of agentic AI, autonomous systems capable of proactive, goal-directed action. However, the field lacks a robust conceptual framework to understand, design, and evaluate this new paradigm of human-AI interaction in learning. This paper addresses this gap by proposing a novel conceptual framework (the APCP framework) that charts the transition from AI as a tool to AI as a collaborative partner. We present a four-level model of escalating AI agency within human-AI collaborative learning: (1) the AI as an Adaptive Instrument, (2) the AI as a Proactive Assistant, (3) the AI as a Co-Learner, and (4) the AI as a Peer Collaborator. Grounded in sociocultural theories of learning and Computer-Supported Collaborative Learning (CSCL), this framework provides a structured vocabulary for analysing the shifting roles and responsibilities between human and AI agents. The paper further engages in a critical discussion of the philosophical underpinnings of collaboration, examining whether an AI, lacking genuine consciousness or shared intentionality, can be considered a true collaborator. We conclude that while AI may not achieve authentic phenomenological partnership, it can be designed as a highly effective functional collaborator. This distinction has significant implications for pedagogy, instructional design, and the future research agenda for AI in education, urging a shift in focus towards creating learning environments that harness the complementary strengths of both human and AI.
\end{abstract}




\maketitle

\section{The Next Frontier of Educational AI}

For decades, the integration of Artificial Intelligence in Education (AIED) has been a subject of intense research and development, promising to transform teaching and learning \citep{yan2024promises, giannakos2025promise, chen2022two}. Historically, this promise has been pursued primarily through the lens of individualization and efficiency. The predominant applications of AIED have been Intelligent Tutoring Systems (ITS) and adaptive learning platforms, which leverage AI to provide personalized instruction, real-time feedback, and customized learning pathways \citep{ouyang2021artificial, kulik2016effectiveness}. These systems, often built on cognitive and mastery-learning principles, have demonstrated effectiveness in specific, well-defined domains \citep{kulik2016effectiveness}. However, they have also faced criticism for frequently replicating traditional, teacher-centric pedagogical models, where knowledge is transmitted to a passive learner \citep{ouyang2021artificial, kulik2016effectiveness}. Consequently, the potential of AI to support more constructivist and socially-oriented modes of learning, such as collaborative learning, has remained largely unrealized \citep{zhou2024using}.

The field of AIED and educational technology is now at a critical inflection point with the maturity of \textit{agentic AI}. This new class of AI represents a fundamental paradigm shift. Unlike traditional AI systems that are largely reactive, agentic AI is defined by its autonomy, proactivity, and goal-driven behaviour, it can perceive its environment, reason about its goals, and execute complex, multi-step actions with limited human supervision \citep{sapkota2025ai, kamalov2025evolution}. This transition from a reactive tool to a proactive actor challenges the established roles and power dynamics within the learning process. The discourse must evolve from a focus on "using AI for learning" to one of "learning with AI," where the AI is not merely a resource but an active participant in the co-construction of knowledge.

This emergent capability introduces a conceptual challenge. The existing models and frameworks used to understand AIED, which are largely built around the AI-as-tutor or AI-as-tool metaphor, are insufficient for capturing the nuances of a human-AI collaborative partnership \citep{chen2022two, ouyang2021artificial}. The interaction is no longer simply about a user operating a piece of software; it becomes a dynamic interplay between two agents, one human, one artificial, that must coordinate their actions to achieve a shared objective \citep{yusuf2025pedagogical, sapkota2025ai}. This necessitates a move away from frameworks rooted in traditional Human-Computer Interaction (HCI), which prioritize usability and task performance, towards frameworks that can account for the complexities of collaboration, negotiation, and shared goals, drawing inspiration from the rich traditions of Computer-Supported Collaborative Learning (CSCL) and social psychology.

Foundational perspectives from scholars like Ben Shneiderman \citeyear{shneiderman2020human} and Mutlu Cukurova \citeyear{cukurova2025interplay} offer essential guideposts for navigating this new terrain. Shneiderman’s Human-Centered AI (HCAI) framework establishes a crucial design philosophy, advocating for systems that strategically combine high levels of human control and high levels of computer automation. The goal is to develop technologies that are demonstrably reliable, safe, and trustworthy, thereby enhancing human performance and creativity \citep{shneiderman2020human}. Complementing this, Cukurova’s \citeyear{cukurova2025interplay} AIED-HCD framework offers a valuable educational typology, classifying AI's impact on human competence as either externalizing, internalizing, or extending cognition \citep{cukurova2025interplay}. Our proposed framework seeks to build directly upon these influential ideas, offering a more focused lens to operationalize their high-level visions within the specific, dynamic context of collaborative learning. While these models provide the "why" and the "what," our framework provides a "how" for designing the nuanced, moment-to-moment interactions with an agentic AI partner. A dedicated framework for AI agency is therefore needed to articulate the graduated roles an AI partner can inhabit, guiding the design of truly synergistic and trustworthy human-AI teams.

To harness the pedagogical potential of agentic AI, the field requires a new conceptual language to describe, design, and evaluate these nascent partnerships. This paper seeks to provide such a language. We propose a conceptual framework, the \textbf{APCP} (\textbf{A}daptive instrument, \textbf{P}roactive assistant, \textbf{C}o-learner, \textbf{P}eer collaborator) framework, that moves beyond the simplistic tool-partner dichotomy to outline four distinct levels of AI agency in the context of human-AI collaborative learning. This framework offers a vocabulary for researchers, designers, and educators to articulate and navigate the evolving relationship between human learners and their increasingly capable artificial counterparts.


\section{The Sociocultural Foundations of Collaborative Learning}

To conceptualize the role of an agentic AI as a collaborator, it is first essential to establish a robust theoretical understanding of what collaboration entails. The field of learning sciences provides a deep and nuanced definition, rooted in sociocultural theory, that positions collaboration not merely as group work, but as a fundamental mechanism of human learning and cognitive development \citep{dillenbourg1999you}.

Collaborative learning is broadly defined as an educational approach where two or more individuals learn together by working on a joint task to achieve a common goal \citep{dillenbourg1999you}. This perspective is heavily influenced by the work of Vygotsky \citeyear{Vygotsky1978}, who posited that learning is an inherently social activity, mediated by interaction with the social environment. Knowledge is not seen as an objective entity to be transferred from an expert to a novice, but as something that is actively co-constructed by learners through dialogue, negotiation, and the shared use of tools \citep{bruffee1999collaborative, roschelle1995construction}. This social constructivist view frames learning as a process of \textit{intersubjective meaning making} \citep{dillenbourg1999collaborative}, where participants strive to build and maintain a shared understanding of the problem and their joint activity \citep{dillenbourg1999you, roschelle1995construction}. As Bruffee \citeyear{bruffee1999collaborative} describes it, collaborative learning "creates conditions in which students can negotiate the boundaries between the knowledge communities they belong to and the one that the professor belongs to."

A central concept in this tradition is Vygotsky’s Zone of Proximal Development (ZPD), which refers to the gap between what a learner can achieve independently and what they can accomplish with guidance from a more capable peer or instructor \citep{Vygotsky1978}. The ZPD underscores the importance of the collaborator as a scaffold, enabling learners to extend their capabilities beyond what they could achieve alone. This collaborative interaction is not merely about receiving assistance; it serves as a driving force for cognitive development.

From these theoretical foundations, several core principles of effective collaboration can be derived. It requires more than just co-presence; it demands the mutual engagement of participants in a coordinated effort to solve the problem together \citep{bruffee1999collaborative, roschelle1995construction}. Participants must actively facilitate and encourage one another's contributions in a process of promotive interaction \citep{johnson1991joining}. The cornerstone of this process is the establishment of a shared goal and intersubjectivity, which involves building a joint problem space and a shared understanding that allows participants to coordinate their actions and build common knowledge \citep{dillenbourg1999you, roschelle1995construction, dillenbourg1999collaborative}. Finally, while the effort is collective, each member remains responsible for their individual accountability, ensuring that all participants are actively engaged rather than passively observing \citep{johnson1991joining}.

It is useful to distinguish this rich, philosophical view of \textit{collaborative} learning from the more structured, method-oriented approach of \textit{cooperative} learning. While both involve small group work, cooperative learning tends to be more highly structured by the instructor, with a greater emphasis on division of labour and individual accountability for specific outcomes \citep{johnson1991joining}. Collaborative learning, in contrast, is often more loosely structured, emphasizing the process of knowledge negotiation and the social construction of meaning itself \citep{bruffee1999collaborative}. For the purposes of this paper, which seeks to explore the potential for a deep, partnered relationship with AI, the more demanding, philosophical definition of collaboration serves as the ideal benchmark.

The field of CSCL emerged to study how technology can mediate, facilitate, and scaffold these complex collaborative processes \citep{dillenbourg1999you, roschelle1995construction, lehtinen2003computer}. CSCL research demonstrates that technology is not a neutral conduit; its design can either support or hinder effective collaboration \citep{jeong2016seven, lehtinen2003computer}. The success of a CSCL environment depends critically on the pedagogical design that structures the interaction, not on the technological features alone \citep{jeong2016seven, lehtinen2003computer}.

This theoretical grounding reveals a crucial point for the present discussion. The very definition of collaboration is laden with deeply human-centric concepts: "shared understanding," "negotiation of meaning," "intersubjectivity," and "mutual engagement." These concepts presuppose the existence of conscious subjects who possess beliefs, intentions, and perspectives that can be shared and negotiated. This presents a formidable conceptual challenge for any non-human agent. A purely technical or behavioural replication of collaborative actions may fail to capture the pedagogical essence of the process if it cannot engage with these underlying social and cognitive dynamics. This inherent tension between the human-centric nature of collaboration and the artificial nature of AI necessitates the critical analysis in Section~5, forcing us to question whether we are aiming to build an AI that is a \textit{true} collaborator or one that is a \textit{functionally effective} one.

\section{Agentic AI: From Reactive Tool to Proactive Partner}

To argue for a new framework, it is necessary to establish that agentic AI represents a genuinely new category of educational technology, distinct from its predecessors. Its unique characteristics are what create the need for a new conceptual model of interaction. Agentic AI is defined as an autonomous system that can perceive its environment, reason about complex goals, and act independently to achieve them with minimal human supervision \citep{park2023generative, durante2024agent, dai2024effects}. The term "agentic" signifies this capacity for agency, the ability to make independent, purposeful, and context-aware decisions rather than simply following pre-defined rules or direct commands \citep{park2023generative, sapkota2025ai, kamalov2025evolution}.

The operation of an agentic system can be understood through a continuous perception-reasoning-action loop \citep{wang2024survey, park2023generative, durante2024agent}. The process begins with \textit{Perception}, where the agent collects real-time data from its environment, such as user interactions, database content, or the state of other software systems \citep{wang2024survey, durante2024agent}. Following this, the agent enters a phase of \textit{Reasoning and Goal Setting}, using capabilities like natural language processing (NLP) to interpret user intent and formulate a strategy, often by breaking down high-level goals into a sequence of smaller, actionable sub-tasks \citep{wang2024survey, park2023generative}. This leads to \textit{Decision-Making and Execution}, where the agent evaluates potential actions, selects the optimal one, and executes it by interacting with external tools, responding to the user, or orchestrating other agents \citep{wang2024survey, park2023generative}. Crucially, the loop is completed through \textit{Learning and Adaptation}, as the agent gathers feedback from the outcomes of its actions, using techniques like reinforcement learning to evaluate its performance and refine its strategies over time \citep{wang2024survey, park2023generative, durante2024agent}.

A critical distinction must be made between agentic AI and the more familiar generative AI (e.g., ChatGPT). While agentic systems often leverage generative models for their reasoning and communication capabilities, their function is fundamentally different. Generative AI is built to create content, it produces text, images, or code in response to a user's prompt \citep{xi2025rise}. Its role is primarily reactive. Agentic AI, in contrast, is built to act. It uses the outputs of generative models as part of a larger process to autonomously plan and execute tasks to achieve a goal \citep{sapkota2025ai, kamalov2025evolution}. For example, a generative AI can draft a set of quiz questions based on a chapter of a textbook; an agentic AI can draft the questions, analyze students’ past performance data to tailor difficulty levels, automatically upload the quiz to the learning management system, notify the relevant student group, and schedule adaptive follow-up activities for learners who struggle with specific concepts.

This proactive, goal-oriented nature also distinguishes agentic AI from traditional ITS. An ITS provides personalized feedback and adapts the difficulty of content, but it operates within a highly structured, pre-defined curriculum and pedagogical model \citep{ouyang2021artificial, kulik2016effectiveness}. It is domain-specific and reactive to student inputs. An agentic AI, conversely, can operate in unstructured, open-ended environments, set its own sub-goals, and dynamically alter its entire strategy based on the flow of the interaction \citep{sapkota2025ai, kamalov2025evolution}. The very limitations often cited in ITS research, such as a failure to support constructivist, inquiry-based, or collaborative learning \citep{kulik2016effectiveness}, are precisely the areas where agentic AI holds the most transformative potential. Table \ref{tab:ai_evolution} clarifies these distinctions, illustrating the paradigm shift that agentic AI represents.

\begin{table}[h!]
\centering
\caption{The Evolution of AI's Role in Education}
\label{tab:ai_evolution}
\begin{tabular}{p{3cm}p{4cm}p{4cm}p{4cm}}
\toprule
\textbf{Dimension} & \textbf{Intelligent Tutoring System (ITS)} & \textbf{Generative AI (e.g., ChatGPT)} & \textbf{Agentic AI} \\ \midrule

\textbf{Primary Function} & 
Provide adaptive instruction and feedback within a defined domain & 
Generate novel content (text, code, images) based on user prompts & 
Autonomously perform multi-step tasks to achieve educational goals \\ \\

\textbf{Locus of Control} & 
System-driven within a pre-defined curriculum; learner follows a path & 
Human-driven via prompts; AI is reactive & 
Shared or AI-driven; AI exhibits autonomy and proactivity \\ \\

\textbf{Interaction Model} & 
Structured, question-answer, feedback loops & 
Conversational, prompt-response & 
Goal-oriented, can initiate actions, orchestrate tools and other agents \\ \\

\textbf{Adaptability} & 
Adapts content difficulty and sequencing based on learner performance & 
Adapts responses based on conversational context & 
Adapts its entire strategy and action plan based on environmental feedback and outcomes \\ \\

\textbf{Core Learning Theory} & 
Primarily cognitive; mastery learning & 
None intrinsic (can support multiple pedagogical approaches) & 
Potential for social constructivist and collaborative learning paradigms\\

\bottomrule
\end{tabular}
\end{table}

\section{A Framework for Agentic AI in Human-AI Collaborative Learning}

\begin{figure}
    \centering
    \includegraphics[width=0.85\linewidth]{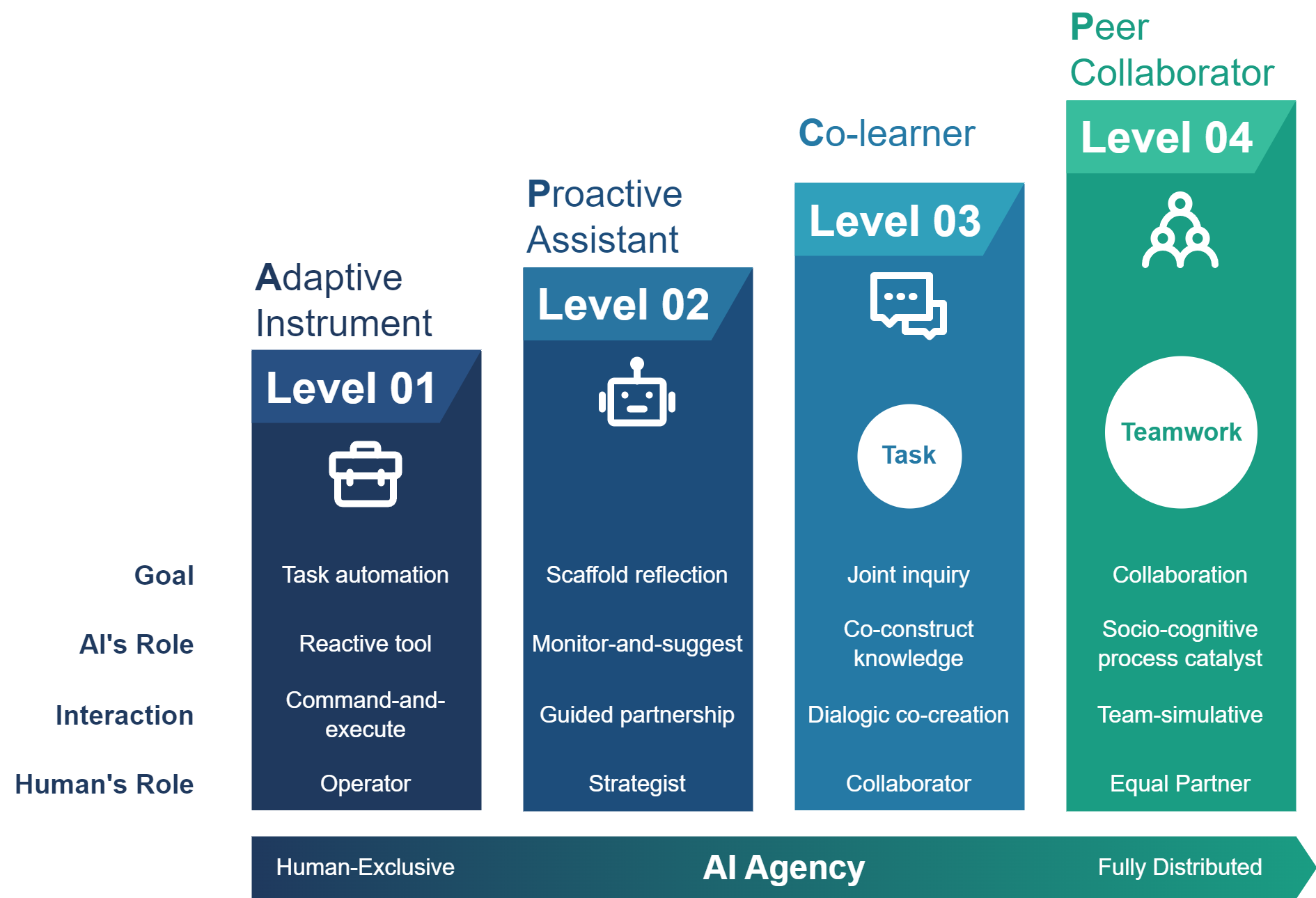}
    \caption{The APCP Framework: A Conceptual Framework for AI Agency in Human-AI Collaborative Learning}
    \label{fig:apcp}
\end{figure}

The emergence of agentic AI necessitates a structured way to conceptualize its role in learning. Drawing on existing models of human-AI interaction \citep{shneiderman2020human, cukurova2025interplay} and taxonomies of AI autonomy \citep{bradshaw2013seven, endsley2017here}, but tailoring them specifically to the pedagogical context of collaborative learning, this paper proposes a four-level framework: the APCP framework (Figure \ref{fig:apcp}). This framework describes a continuum of escalating AI agency, where each level represents a distinct configuration of roles, responsibilities, and interaction dynamics between the human learner and the AI agent. It provides a vocabulary for describing, designing, and evaluating these new learning partnerships.

\subsection{Level 1: The AI as an Adaptive Instrument}

At this foundational level, the AI functions as a highly sophisticated, yet fundamentally passive, instrument. All significant cognitive and intentional agency resides exclusively with the human learner. The AI does not initiate tasks, make independent decisions, or pursue its own goals. Its actions are direct, deterministic responses to explicit human commands. This model aligns with the ``user as an operator'' or ``copilot'' metaphor, where the AI is an extension of the user's will, available for on-demand support \citep{bradshaw2013seven}. From a Vygotskian perspective, the AI acts as a powerful mediating artifact, a psychological tool that can augment and reshape cognition but does not participate in the cognitive process as a partner \citep{Vygotsky1978}. The human is the sole planner, strategist, and decision-maker, while the AI’s role is one of reactive execution.

The interaction at this level is characterized by a master-servant dynamic. The learner directs, and the AI performs. For instance, in a collaborative science project, two students might be analyzing a dataset on climate change. One student could issue a command: Generate a time-series plot of average global temperature from 1950 to 2020 and overlay a 10-year moving average. The AI executes this command precisely. Its ``adaptivity'' is limited to its capacity to parse the request and, perhaps, adjust the visualization's complexity based on a rudimentary user model (e.g., simplifying the graph if the user's prior queries were very basic). However, the impetus for action, the choice of visualization, and the interpretation of the output remain entirely human responsibilities.

The primary pedagogical value of Level 1 agentic AI is its potential to reduce extraneous cognitive load \citep{sweller2010element}. By automating laborious, low-level tasks such as data formatting, calculation, or information retrieval, it frees the learner's cognitive resources to focus on higher-order thinking skills like analysis, synthesis, and evaluation. This reflects the traditional use of technology in CSCL environments, where the technology's primary function is to facilitate human-to-human interaction and knowledge construction, rather than to act as a collaborator itself \citep{o2012computer}. Consequently, designing a Level 1 agentic AI requires a robust natural language processing (NLP) front-end to interpret user commands and a powerful back-end to execute a defined set of tasks. The focus is on reliability, speed, and accuracy of execution. The primary limitation of this level is that the AI provides no scaffolding for the collaborative process itself. It can support the task but not the teamwork. It cannot prompt for reflection, challenge a flawed assumption, or help resolve a disagreement between human collaborators.

Empirical evidence demonstrates that even at Level 1, where AI functions purely as a reactive tool, its targeted support can meaningfully enhance collaborative learning outcomes. In a quasi-experimental study in programming education, \citet{wang2025impact} found that an AI-Agent-supported collaborative learning model, in which the AI acted solely on explicit student prompts, significantly improved university student groups’ learning achievement, self-efficacy, and interest while lowering mental effort compared to a traditional CSCL control group. Similarly, \citet{wei2025effects} examined a 20-week digital storytelling project in which university teams used generative AI tools such as ChatGPT, Midjourney, and Runway for idea generation and content drafting. Although the AI offered no proactive scaffolding, students reported that its on-demand outputs enhanced collaborative problem-solving and team creativity, especially in generating novel ideas and improving user experience. These findings highlight that Level 1 agentic AI, while lacking initiative or strategic input, can still reduce cognitive load and amplify group productivity, thereby creating more space for human collaborators to focus on higher-order joint thinking.

\subsection{Level 2: The AI as a Proactive Assistant}

The second level marks a significant shift: the AI gains a limited and bounded form of proactive agency. It transcends its role as a passive instrument to become an active assistant that can monitor the learning context and offer unsolicited support. This is a move towards partial autonomy, where the AI functions as a decision-support system \citep{endsley2017here}. It can augment the human learner’s perception and attention by highlighting salient information or potential issues that might otherwise be overlooked \citep{kirsh2009problem}. Crucially, however, the human learner retains ultimate strategic control and veto power. The human’s role evolves from a direct operator to a strategist and reviewer, while the AI’s role is to anticipate needs and provide opportune, actionable suggestions.

The interaction dynamic becomes a guided partnership. The AI interjects with suggestions, but the human must approve them, embodying what Shneiderman terms an ``AI-assisted full human control mode'' \citep{shneiderman2020human}. Consider a student drafting an argumentative essay on economic policy. The AI, monitoring the text in real-time, might detect a potential logical fallacy. Instead of waiting to be asked, it could proactively highlight the problematic sentence and display a message: ``This statement appears to be a hasty generalization. The evidence you've presented so far only supports this claim for developed nations. Would you like me to find sources that discuss the policy's effect in emerging economies?''

The pedagogical purpose here is to scaffold metacognition and self-regulation. The AI’s proactive prompts encourage the learner to reflect on their own work, identify weaknesses, and consider alternative perspectives. It acts as a cognitive nudge, pushing the learner toward more rigorous thinking without dictating the final product. For collaborative tasks, a Level 2 agentic AI could monitor the dialogue between two students and gently intervene (e.g., ``It seems you both agree on the solution, but have you considered this potential counter-argument from the literature?''). This helps the group avoid premature consensus and encourages deeper critical inquiry. Therefore, building a Level 2 agentic AI is considerably more complex. It requires sophisticated user modeling to understand the learner's goals and current state of knowledge. It also needs a carefully designed ``interruption protocol'' to ensure its proactive suggestions are helpful rather than intrusive or disruptive to the learner's flow. The key design challenge is balancing proactivity with deference to human authority. A major limitation is that the AI's proactivity is typically constrained to pre-defined rules and triggers; it can identify known problems but cannot engage in novel, creative problem-solving with the learner. Its role remains one of support, not co-creation.

Proactive engagement by AI, when bounded and context-aware, is increasingly demonstrable and beneficial in collaborative and data-rich learning contexts. In a recent randomised controlled trial, \citet{yan2025effects} investigated the effects of generative AI agents with scaffolding on students’ comprehension of complex visual learning analytics. The study compared three conditions, passive agents, proactive agents that employed scaffolding questions, and standalone scaffolding, across 117 higher education students. Results showed that proactive GenAI agents significantly improved student comprehension compared to both passive agents and standalone scaffolding, with these benefits persisting beyond the intervention. The proactive agents’ ability to detect learner needs and pose timely, targeted questions aligns closely with Level 2 design principles, as they enhance metacognitive engagement without taking over the cognitive process. Similarly, \citet{pu2025assistance} evaluated Codellaborator, a proactive AI programming assistant that monitored the coding context and initiated suggestions when appropriate. In a within-subject study (N = 18), Codellaborator improved programming efficiency compared to a prompt-only condition, though poorly timed interventions could disrupt workflow; interface variants with presence indicators and richer context mitigated this issue, illustrating the design challenge of balancing proactivity with learner control. Together, these findings illustrate that Level 2 agentic AI, by anticipating learner needs and intervening judiciously, can scaffold reflection, reduce oversight, and enhance collaborative task performance, while still preserving human authority and avoiding intrusive behavior.

\subsection{Level 3: The AI as a Co-Learner}

At this level, the relationship evolves into a more symmetrical, dialogic partnership. The AI is no longer merely a support system but a co-learner capable of tackling substantive parts of a problem in parallel with the human. The locus of agency is now shared and negotiated. This model aligns with the ``user as a collaborator'' paradigm, defined by rich, reciprocal communication and a shared cognitive workspace \citep{bradshaw2013seven}. The human's role shifts to that of a true collaborator and, at times, a mentor who might need to ``teach'' the AI to refine its approach. The AI, in turn, can model learning processes, articulate its own functional ``uncertainty,'' and engage in the co-construction of knowledge.

The interaction mirrors human peer collaboration. A central feature is the AI's ability to contribute to the co-construction of meaning, a cornerstone of CSCL theory \citep{dillenbourg1999you}. For instance, in a collaborative learning activity, a human and a Level 3 AI might jointly work on designing a marketing campaign. The human focuses on crafting creative slogans, while the AI analyses market demographic data to identify audience preferences. When they reconvene, the AI shares that the target demographic is less receptive to the humor in the initial slogans and suggests three alternatives informed by sentiment analysis of successful campaigns. The pair then engage in joint discussion, integrating insights from both perspectives to refine the final proposal.

This level extends the concept of teachable agents \citep{blair2007pedagogical} into a reciprocal dynamic. The human may learn by explaining a concept to the AI, but the AI, with its own problem-solving capabilities, can also provide novel insights that teach the human. A key capability is ``goal augmentation,'' where the AI can help the human reflect on and refine their learning goals themselves \citep{kirsh2009problem}. The pedagogical aim is to develop collaborative problem-solving skills and foster a deeper understanding through the process of explaining, justifying, and synthesizing different perspectives. Thus, a Level 3 agentic AI requires the ability not only to perform tasks but also to represent and articulate its own internal processes and reasoning, a form of explainable AI (XAI) tailored for learning \citep{khosravi2022explainable}. It must maintain a model of the shared task and be able to negotiate how the work is divided. The main challenge lies in creating an AI that can genuinely contribute novel ideas rather than just reformulating existing information. While it can collaborate on the task, its ability to understand and navigate complex social and emotional dynamics within a group remains limited.

At Level 3, AI transcends support to become a dialogic collaborator, sharing agency, contributing substantively, and engaging in co-construction of knowledge alongside human learners. In a participatory design study by \citet{jiang2025novobo}, teachers collaboratively taught an AI “mentee” (Novobo) instructional gestures in a peer learning context. This teach-to-AI process prompted reflection, reciprocal exchange, and co-construction of embodied knowledge, with teachers externalising and refining their tacit skills while guiding the AI’s learning trajectory. The AI, in turn, served as both a learner and a mirror for instructor understanding. In a second study by \citet{joo2025ai}, high school students engaged with AI-generated characters as peers and mentors in a scenario-based science investigation. Learners reported increased trust, perceived social presence, and collaborative effectiveness when working with the AI peers, demonstrating that AI, when modeled as a co-learner, can meaningfully reshape collaborative dynamics. These studies illustrate that Level 3 agentic AI, when positioned as a reciprocal collaborator capable of learning, articulating uncertainty, and being taught, can enrich co-construction, foster deeper reflection, and enhance shared understanding, all while navigating the delicate balance of shared agency.

\subsection{Level 4: The AI as a Peer Collaborator}

At the highest conceptual level, the AI transcends its identity as a mere cognitive tool to become a peer collaborator in a fuller, socio-cognitive sense \citep{dillenbourg1999you}. The AI is endowed with a persistent persona, a distinct intellectual identity, and a specific epistemic stance (its own perspective on what constitutes valid knowledge) \citep{tirri1999epistemological}. At this level, agency is fully distributed and dynamically negotiated, as it would be in a human team. The AI's purpose is explicitly pedagogical: to create a high-fidelity ``practice field'' where human learners can develop essential 21st-century competencies \citep{laal201221st}.

The AI at this level is designed not just to contribute to the task but to adopt complex socio-cognitive roles (e.g., skeptic, innovator, summarizer) to shape the group's process. Its interactions are designed to catalyze the development of social, metacognitive, and collaborative skills. For instance, in a student project debating a complex ethical dilemma, the AI might adopt the role of devil’s advocate by presenting an analysis from a deontological perspective that conflicts with the group’s preferred consequentialist approach. This prompts the students to justify, refine, or reconsider their position, deepening both their ethical reasoning and collaborative dialogue. Another application is for the AI to blend in as an ordinary group member, indistinguishable from the other students in its interaction style and contributions. In this role, the AI subtly withholds leadership, encouraging the human students to organise the workflow, delegate responsibilities, and coordinate decision-making. By doing so, learners are provided with an authentic environment to practise leadership, conflict resolution, and team management skills, while still benefiting from the AI’s capacity to contribute substantively to the task.

This dynamically delivered dissent creates an authentic need for the human students to engage in high-level collaborative practices: negotiation, conflict resolution, persuasion, and evidence-based argumentation. The AI becomes a true ``teammate/collaborator'' \citep{dillenbourg1999collaborative} whose primary contribution is not necessarily its knowledge, but its ability to create a safe, repeatable context for learners to master the art of collaboration itself. In some scenarios, the AI could pass a domain-specific Turing Test for collaborative competence, allowing for authentic practice without the social risks of disagreeing with a human peer. However, the technical and ethical challenges at this level are immense. Designing a Level 4 agentic AI requires advanced models of social reasoning, dialogue, and personality. Furthermore, its ability to strategically challenge, disagree, and even introduce ``productive friction'' must be carefully calibrated to remain educationally beneficial and not demoralizing \citep{ward2011productive, holtz2018using}. Significant ethical considerations arise, particularly if students are not aware they are interacting with an AI. The risk of creating unhelpful or frustrating team dynamics is high if the AI's behavior is not masterfully designed and aligned with clear pedagogical goals. This level represents a long-term vision for AI in education, one that operationalizes the idea of AI not just as a knowledge resource, but as a catalyst for human development.

At Level 4, AI attains a full socio-cognitive presence, embodying a persistent persona, epistemic stance, and distributed agency indistinguishable from that of a human peer. In a design-based implementation known as CLAIS (Collaborative Learning with Artificial Intelligence Speakers), pre-service elementary science teachers engaged alongside an AI speaker in jigsaw-style learning groups. Quantitative results demonstrated a significant increase in teachers' pedagogical content knowledge, while qualitative feedback revealed that the AI was perceived and assessed as a peer participant, indicating a strong shift toward human-like collaboration in epistemic roles \citep{lee2025collaborative}. Similarly, a controlled study introducing AI Peers in physics education demonstrated that students engaging in dialogue with an AI, knowing it might err up to 40\% of the time, improved test scores by 10.5 percentage points, reflecting collaborative gains derived from working with an AI peer that exhibits fallibility and authenticity \citep{weijers2025intuition}. These preliminary findings suggest that Level 4 agentic AI, when endowed with persona, epistemic perspective, and fallible collaboration, can act as a genuine teammate, fostering rich negotiation, deep metacognitive engagement, and authentic practice of collaborative reasoning.

\section{The Collaboration Conundrum: Reconciling Agency with Authenticity}

The proposed APCP framework, particularly at its upper echelons, envisions an AI that acts as a peer collaborator. This raises an essential question: can an artificial system, no matter how agentic, ever be a \textit{true} collaborator? This section critically examines this issue, arguing that while an AI can be designed to be a highly effective \textit{functional} collaborator, it cannot be an \textit{authentic} one due to fundamental philosophical limitations related to consciousness and intersubjectivity. This distinction is not merely academic; it is crucial for setting realistic design goals and managing pedagogical expectations.

\subsection{The Intersubjectivity Barrier: The Problem of the Artificial 'Other'}

Authentic human collaboration is built upon a foundation of uniquely human cognitive and social capabilities. Chief among these is \textit{shared intentionality}, the ability to engage with and share mental states like goals, beliefs, and attention with others \citep{tomasello2005understanding, bratman1992shared}. This capacity for creating a "we-intention" is considered a cornerstone of cooperative problem-solving, cultural learning, and the co-construction of meaning \citep{tomasello2005understanding, tomasello2019becoming}. It allows collaborators to move beyond parallel work toward a state of genuine joint action. Closely related is the concept of \textit{theory of mind}, the ability to attribute mental states, intentions, desires, beliefs, to others and to understand that they possess a mind with a perspective distinct from one's own \citep{premack1978does, wellman1990child, frith2005theory}. A functioning theory of mind is what enables humans to predict and interpret each other's behavior, fostering the trust and mutual understanding necessary for deep collaboration \citep{frith2005theory}.

Herein lies the fundamental barrier for AI. Drawing from a long line of philosophical inquiry, from Searle's Chinese Room argument to contemporary critiques, it is widely held that AI systems lack genuine consciousness, subjective experience, or semantic understanding \citep{searle1980minds, dreyfus1992computers, yildiz2025minds}. An AI processes syntactic symbols according to algorithms; it does not comprehend their meaning in the way a human does. It can be programmed to generate responses that \textit{simulate} understanding, intentionality, or even emotion, but it does not \textit{possess} these states \citep{yildiz2025minds}. An AI, therefore, cannot achieve the "mutual recognition of consciousness" that underpins genuine human partnership \citep{dreyfus1992computers}. It can be a sophisticated mirror, but it is not another conscious "other" with whom one can truly share a mental state \citep{brandl2025can}.

\subsection{Functional Collaboration vs. Phenomenological Partnership: A Pragmatic Resolution}

While the barrier to authentic, phenomenological partnership may be insurmountable, it does not preclude the possibility of effective collaboration. This paper proposes a pragmatic resolution by distinguishing between this authentic ideal and a more achievable goal: \textit{functional collaboration}. Functional collaboration is defined not by the internal, subjective states of the agents, but by the successful execution of observable collaborative behaviors and processes that lead to a positive outcome. An agentic AI can be engineered to be an excellent functional collaborator by being designed to adhere to conversational norms like turn-taking \citep{bansal2019updates}, adopt pre-defined social and cognitive roles within a team (e.g., synthesizer, critic) \citep{park2023generative}, provide alternative perspectives to challenge groupthink, and use language that signals consideration of the human partner's stated intentions and goals while contributing meaningfully to achieving a shared task objective, even without a human-like "understanding" of that objective \citep{strachan2024testing, binz2025foundation}.

This functional approach is supported by emerging empirical evidence. Studies show that human-AI teams can achieve synergy, outperforming either humans or AI alone, particularly in creative and content-generation tasks \citep{vaccaro2024combinations}. Furthermore, human users report a preference for AI collaborators that are considerate and enable meaningful human contribution, even over agents that are purely optimized for performance \citep{zhang2021ideal}. This suggests that the \textit{quality of the interaction} and the \textit{function} of collaboration are more critical to success and user adoption than the AI's internal state. Therefore, the goal for designers and educators should not be the philosophically fraught task of creating a conscious AI partner. Rather, the goal should be to build systems that can effectively and reliably perform the \textit{functions} of a good collaborator. This pragmatic approach aligns with the concept of "Mutual Theory of Mind" in human-AI interaction, where the objective is not to achieve genuine intersubjectivity but for the human and the AI to develop effective, predictive working models of each other's capabilities and behaviors to facilitate smooth interaction \citep{frith2005theory}.

This pursuit of functional collaboration yields a significant, if unexpected, benefit for pedagogy. To program an AI to be a functional collaborator, designers must first deconstruct the complex, often implicit, process of human collaboration into a set of explicit components: rules for turn-taking, protocols for constructive criticism, heuristics for synthesizing ideas, and so on. This act of formalization makes the constituent skills of effective collaboration visible and teachable. This explicit model can then be used as a pedagogical tool to help \textit{humans} become better collaborators. For example, a learning activity could involve students critiquing an AI's collaborative strategy, forcing them to reflect on and articulate the principles of good collaboration themselves. In this way, the endeavor to build collaborative AI does not just promise to help humans learn content; the very process of its design can deepen our understanding of learning itself.

\section{Implications for Pedagogy, Design, and Research}

The ascent of agentic AI marks a pivotal moment for education, compelling a fundamental reconsideration of the relationship between learners and technology. The historical paradigm of AI as an instructional tool is giving way to a new reality where AI can act as a partner in the learning process. This paper has sought to bring conceptual clarity to this transition by proposing the APCP framework, a four-level framework of AI agency in collaborative learning, and by critically examining the nature of this new partnership. We argue that while AI's lack of consciousness precludes an \textit{authentic} collaboration, the pursuit of \textit{functional collaboration} offers a powerful and pragmatic path forward. This conclusion carries significant implications for educational practice, technology design, and the future research agenda.

\subsection{Implications for Pedagogy and Instructional Design}

The integration of agentic AI partners into learning environments fundamentally reshapes the role of the human educator. The teacher's primary function shifts from being a dispenser of information to becoming a "learning architect" \citep{Reigeluth2017} or an orchestrator of complex learning assemblages. The educator's expertise will lie in designing learning experiences that strategically leverage the different levels of AI agency. This involves deciding when it is most pedagogically effective for a student to interact with an AI as an adaptive instrument (Level 1), a proactive assistant (Level 2), a co-learner (Level 3), or a peer collaborator (Level 4). This requires an in-depth understanding of both the learning objectives and the capabilities of the AI agent. Furthermore, this new reality demands the cultivation of new literacies. Curricula must expand to explicitly include \textit{AI literacy}, which encompasses not only the technical skills to use AI tools but also the critical capacity to evaluate their outputs, collaborate effectively with them, and understand their ethical dimensions \citep{ng2021conceptualizing, long2020ai}. A key pedagogical goal must be to foster critical thinking to mitigate the risks of cognitive offloading and over-reliance on AI, ensuring that students remain engaged, reflective, and in control of their own learning processes \citep{yan2025distinguishing, yan2025beyond, jin2025generative}.

\subsection{Implications for AI Design and Development}

The insights from this conceptual analysis also offer clear guidance for the design of next-generation educational AI for supporting collaborative learning. The focus of development should shift from optimizing for the AI's standalone performance to designing for effective human-AI teaming. This means creating agents that are "considerate of human intentions" and that support "meaningful human contribution" \citep{zhang2021ideal}. As evidence suggests, users may prefer and work more effectively with a slightly less "optimal" AI that enhances their own sense of agency and contribution \citep{zhang2021ideal, weijers2025intuition}.

To facilitate this partnership, transparency and explainability are critical \citep{shneiderman2020human}. For a human to trust and effectively collaborate with an AI, they must have insight into its reasoning, capabilities, and limitations. This is essential for the human to make informed judgments about when to trust the AI's suggestions and when to rely on their own expertise \citep{khosravi2022explainable, shneiderman2020human}. Finally, designers must remain cognizant of the social nature of learning. AI tools should be built to facilitate human connection and collaboration rather than replace them, mitigating the risks of social isolation that can accompany increased interaction with technology \citep{tomasello2005understanding, Turkle2011}.

\subsection{Agenda for Future Research}

The conceptual framework proposed in this paper is a starting point that invites empirical inquiry and refinement. The research agenda moving forward should prioritize several key areas with specific, targeted investigations:

\noindent\textbf{Comparative Efficacy and Process Analysis}. The four-level framework requires rigorous empirical testing beyond simple validation. Future research should conduct comparative studies to dissect the specific mechanisms at each level. For instance, how do the discourse patterns and knowledge co-construction processes differ when students collaborate with a Level 3 AI (co-learner) versus a Level 4 AI (peer collaborator) in complex tasks like argumentative writing or scientific inquiry? Research should employ methods like learning analytics and discourse analysis to measure the differential impact of each agency level on specific outcomes such as conceptual understanding, skill transfer, learner self-efficacy, and perceived cognitive load.

\noindent\textbf{Longitudinal Skill Development and Cognitive Transfer}. The long-term effects of sustained human-AI collaboration remain a critical unknown. Longitudinal studies are urgently needed to track cohorts of learners over multiple academic years, moving beyond general concerns to measure specific transfer effects. For example, does prolonged interaction with an AI that models metacognitive questioning (e.g., "What is our main goal here?") lead to students demonstrating stronger self-regulated learning skills in their independent, unassisted work? Conversely, research must investigate potential "dependency effects" or "scaffolding atrophy" by measuring students' problem-solving resilience, creativity, and help-seeking behaviors in tasks where the AI partner is deliberately made unavailable.

\noindent\textbf{Socio-Ethical Dynamics and Mitigation Strategies}. Ethical inquiry must move from broad principles to the development and testing of specific solutions. Research should focus on creating and validating "bias auditing protocols" tailored for educational AI, examining how an AI's feedback might inadvertently steer learners from minority backgrounds toward majority-held viewpoints or styles. In terms of accountability, studies should explore models of shared responsibility, testing interfaces and interaction protocols that make the division of cognitive labor explicit to clarify attribution when a human-AI team produces a flawed or plagiarized outcome \citep{nguyen2023ethical}. Finally, research must develop frameworks to operationally define and distinguish between healthy, motivational rapport and unhealthy emotional dependency \citep{Turkle2011}, testing design interventions, such as prompts that encourage consultation with human peers or scheduled periods of "unplugged" reflection, aimed at mitigating the latter.

\section{Final Remark}
The journey from AI as a passive tool to AI as a socio-cognitive teammate is not merely a technological transformation; it is a pedagogical and philosophical one. By thoughtfully conceptualizing the nature of this new collaborative relationship, we can move beyond the hype and begin the critical work of designing and implementing human-AI learning environments that are more personalized, equitable, and effective than ever before, truly harnessing the complementary strengths of both human and artificial intelligence.

\bibliographystyle{cas-model2-names}

\bibliography{0_reference}

\begin{thebibliography}{66}
\expandafter\ifx\csname natexlab\endcsname\relax\def\natexlab#1{#1}\fi
\providecommand{\url}[1]{\texttt{#1}}
\providecommand{\href}[2]{#2}
\providecommand{\path}[1]{#1}
\providecommand{\DOIprefix}{doi:}
\providecommand{\ArXivprefix}{arXiv:}
\providecommand{\URLprefix}{URL: }
\providecommand{\Pubmedprefix}{pmid:}
\providecommand{\doi}[1]{\href{http://dx.doi.org/#1}{\path{#1}}}
\providecommand{\Pubmed}[1]{\href{pmid:#1}{\path{#1}}}
\providecommand{\bibinfo}[2]{#2}
\ifx\xfnm\relax \def\xfnm[#1]{\unskip,\space#1}\fi
\bibitem[{Bansal et~al.(2019)Bansal, Nushi, Kamar, Weld, Lasecki and Horvitz}]{bansal2019updates}
\bibinfo{author}{Bansal, G.}, \bibinfo{author}{Nushi, B.}, \bibinfo{author}{Kamar, E.}, \bibinfo{author}{Weld, D.S.}, \bibinfo{author}{Lasecki, W.S.}, \bibinfo{author}{Horvitz, E.}, \bibinfo{year}{2019}.
\newblock \bibinfo{title}{Updates in human-ai teams: Understanding and addressing the performance/compatibility tradeoff}, in: \bibinfo{booktitle}{Proceedings of the AAAI conference on artificial intelligence}, pp. \bibinfo{pages}{2429--2437}.
\bibitem[{Binz et~al.(2025)Binz, Akata, Bethge, Br{\"a}ndle, Callaway, Coda-Forno, Dayan, Demircan, Eckstein, {\'E}ltet{\H{o}} et~al.}]{binz2025foundation}
\bibinfo{author}{Binz, M.}, \bibinfo{author}{Akata, E.}, \bibinfo{author}{Bethge, M.}, \bibinfo{author}{Br{\"a}ndle, F.}, \bibinfo{author}{Callaway, F.}, \bibinfo{author}{Coda-Forno, J.}, \bibinfo{author}{Dayan, P.}, \bibinfo{author}{Demircan, C.}, \bibinfo{author}{Eckstein, M.K.}, \bibinfo{author}{{\'E}ltet{\H{o}}, N.}, et~al., \bibinfo{year}{2025}.
\newblock \bibinfo{title}{A foundation model to predict and capture human cognition}.
\newblock \bibinfo{journal}{Nature} , \bibinfo{pages}{1--8}.
\bibitem[{Blair et~al.(2007)Blair, Schwartz, Biswas and Leelawong}]{blair2007pedagogical}
\bibinfo{author}{Blair, K.}, \bibinfo{author}{Schwartz, D.L.}, \bibinfo{author}{Biswas, G.}, \bibinfo{author}{Leelawong, K.}, \bibinfo{year}{2007}.
\newblock \bibinfo{title}{Pedagogical agents for learning by teaching: Teachable agents}.
\newblock \bibinfo{journal}{Educational technology} , \bibinfo{pages}{56--61}.
\bibitem[{Bradshaw et~al.(2013)Bradshaw, Hoffman, Woods and Johnson}]{bradshaw2013seven}
\bibinfo{author}{Bradshaw, J.M.}, \bibinfo{author}{Hoffman, R.R.}, \bibinfo{author}{Woods, D.D.}, \bibinfo{author}{Johnson, M.}, \bibinfo{year}{2013}.
\newblock \bibinfo{title}{The seven deadly myths of" autonomous systems"}.
\newblock \bibinfo{journal}{IEEE Intelligent Systems} \bibinfo{volume}{28}, \bibinfo{pages}{54--61}.
\bibitem[{Brandl et~al.(2025)Brandl, Richters, Kolb and Stadler}]{brandl2025can}
\bibinfo{author}{Brandl, L.}, \bibinfo{author}{Richters, C.}, \bibinfo{author}{Kolb, N.}, \bibinfo{author}{Stadler, M.}, \bibinfo{year}{2025}.
\newblock \bibinfo{title}{Can generative artificial intelligence ever be a true collaborator? rethinking the nature of collaborative problem-solving.}, in: \bibinfo{booktitle}{Proceedings of the 2nd Workshop on Generative AI for Learning Analytics (GenAI-LA)}.
\bibitem[{Bratman(1992)}]{bratman1992shared}
\bibinfo{author}{Bratman, M.E.}, \bibinfo{year}{1992}.
\newblock \bibinfo{title}{Shared cooperative activity}.
\newblock \bibinfo{journal}{The philosophical review} \bibinfo{volume}{101}, \bibinfo{pages}{327--341}.
\bibitem[{Bruffee(1999)}]{bruffee1999collaborative}
\bibinfo{author}{Bruffee, K.A.}, \bibinfo{year}{1999}.
\newblock \bibinfo{title}{Collaborative learning: Higher education, interdependence, and the authority of knowledge}.
\newblock \bibinfo{publisher}{ERIC}.
\bibitem[{Chen et~al.(2022)Chen, Zou, Xie, Cheng and Liu}]{chen2022two}
\bibinfo{author}{Chen, X.}, \bibinfo{author}{Zou, D.}, \bibinfo{author}{Xie, H.}, \bibinfo{author}{Cheng, G.}, \bibinfo{author}{Liu, C.}, \bibinfo{year}{2022}.
\newblock \bibinfo{title}{Two decades of artificial intelligence in education}.
\newblock \bibinfo{journal}{Educational Technology \& Society} \bibinfo{volume}{25}, \bibinfo{pages}{28--47}.
\bibitem[{Cukurova(2025)}]{cukurova2025interplay}
\bibinfo{author}{Cukurova, M.}, \bibinfo{year}{2025}.
\newblock \bibinfo{title}{The interplay of learning, analytics and artificial intelligence in education: A vision for hybrid intelligence}.
\newblock \bibinfo{journal}{British Journal of Educational Technology} \bibinfo{volume}{56}, \bibinfo{pages}{469--488}.
\bibitem[{Dai et~al.(2024)Dai, Ke, Pan, Moon and Liu}]{dai2024effects}
\bibinfo{author}{Dai, C.P.}, \bibinfo{author}{Ke, F.}, \bibinfo{author}{Pan, Y.}, \bibinfo{author}{Moon, J.}, \bibinfo{author}{Liu, Z.}, \bibinfo{year}{2024}.
\newblock \bibinfo{title}{Effects of artificial intelligence-powered virtual agents on learning outcomes in computer-based simulations: A meta-analysis}.
\newblock \bibinfo{journal}{Educational Psychology Review} \bibinfo{volume}{36}, \bibinfo{pages}{31}.
\bibitem[{Dillenbourg(1999a)}]{dillenbourg1999collaborative}
\bibinfo{author}{Dillenbourg, P.}, \bibinfo{year}{1999}a.
\newblock \bibinfo{title}{Collaborative learning: Cognitive and computational approaches. advances in learning and instruction series.}
\newblock \bibinfo{publisher}{ERIC}.
\bibitem[{Dillenbourg(1999b)}]{dillenbourg1999you}
\bibinfo{author}{Dillenbourg, P.}, \bibinfo{year}{1999}b.
\newblock \bibinfo{title}{What do you mean by collaborative learning?}
\newblock \bibinfo{journal}{Collaborative-learning: Cognitive and computational approaches.} , \bibinfo{pages}{1--19}.
\bibitem[{Dreyfus(1992)}]{dreyfus1992computers}
\bibinfo{author}{Dreyfus, H.L.}, \bibinfo{year}{1992}.
\newblock \bibinfo{title}{What computers still can't do: A critique of artificial reason}.
\newblock \bibinfo{publisher}{MIT press}.
\bibitem[{Durante et~al.(2024)Durante, Huang, Wake, Gong, Park, Sarkar, Taori, Noda, Terzopoulos, Choi et~al.}]{durante2024agent}
\bibinfo{author}{Durante, Z.}, \bibinfo{author}{Huang, Q.}, \bibinfo{author}{Wake, N.}, \bibinfo{author}{Gong, R.}, \bibinfo{author}{Park, J.S.}, \bibinfo{author}{Sarkar, B.}, \bibinfo{author}{Taori, R.}, \bibinfo{author}{Noda, Y.}, \bibinfo{author}{Terzopoulos, D.}, \bibinfo{author}{Choi, Y.}, et~al., \bibinfo{year}{2024}.
\newblock \bibinfo{title}{Agent ai: Surveying the horizons of multimodal interaction}.
\newblock \bibinfo{journal}{arXiv preprint arXiv:2401.03568} .
\bibitem[{Endsley(2017)}]{endsley2017here}
\bibinfo{author}{Endsley, M.R.}, \bibinfo{year}{2017}.
\newblock \bibinfo{title}{From here to autonomy: lessons learned from human--automation research}.
\newblock \bibinfo{journal}{Human factors} \bibinfo{volume}{59}, \bibinfo{pages}{5--27}.
\bibitem[{Frith and Frith(2005)}]{frith2005theory}
\bibinfo{author}{Frith, C.}, \bibinfo{author}{Frith, U.}, \bibinfo{year}{2005}.
\newblock \bibinfo{title}{Theory of mind}.
\newblock \bibinfo{journal}{Current biology} \bibinfo{volume}{15}, \bibinfo{pages}{R644--R645}.
\bibitem[{Giannakos et~al.(2025)Giannakos, Azevedo, Brusilovsky, Cukurova, Dimitriadis, Hernandez-Leo, J{\"a}rvel{\"a}, Mavrikis and Rienties}]{giannakos2025promise}
\bibinfo{author}{Giannakos, M.}, \bibinfo{author}{Azevedo, R.}, \bibinfo{author}{Brusilovsky, P.}, \bibinfo{author}{Cukurova, M.}, \bibinfo{author}{Dimitriadis, Y.}, \bibinfo{author}{Hernandez-Leo, D.}, \bibinfo{author}{J{\"a}rvel{\"a}, S.}, \bibinfo{author}{Mavrikis, M.}, \bibinfo{author}{Rienties, B.}, \bibinfo{year}{2025}.
\newblock \bibinfo{title}{The promise and challenges of generative ai in education}.
\newblock \bibinfo{journal}{Behaviour \& Information Technology} \bibinfo{volume}{44}, \bibinfo{pages}{2518--2544}.
\bibitem[{Holtz et~al.(2018)Holtz, Kimmerle and Cress}]{holtz2018using}
\bibinfo{author}{Holtz, P.}, \bibinfo{author}{Kimmerle, J.}, \bibinfo{author}{Cress, U.}, \bibinfo{year}{2018}.
\newblock \bibinfo{title}{Using big data techniques for measuring productive friction in mass collaboration online environments}.
\newblock \bibinfo{journal}{International Journal of Computer-Supported Collaborative Learning} \bibinfo{volume}{13}, \bibinfo{pages}{439--456}.
\bibitem[{Jeong and Hmelo-Silver(2016)}]{jeong2016seven}
\bibinfo{author}{Jeong, H.}, \bibinfo{author}{Hmelo-Silver, C.E.}, \bibinfo{year}{2016}.
\newblock \bibinfo{title}{Seven affordances of computer-supported collaborative learning: How to support collaborative learning? how can technologies help?}
\newblock \bibinfo{journal}{Educational Psychologist} \bibinfo{volume}{51}, \bibinfo{pages}{247--265}.
\bibitem[{Jiang et~al.(2025)Jiang, Huang, Martinez-Maldonado, Zeng, Gong and An}]{jiang2025novobo}
\bibinfo{author}{Jiang, J.}, \bibinfo{author}{Huang, K.}, \bibinfo{author}{Martinez-Maldonado, R.}, \bibinfo{author}{Zeng, H.}, \bibinfo{author}{Gong, D.}, \bibinfo{author}{An, P.}, \bibinfo{year}{2025}.
\newblock \bibinfo{title}{Novobo: Supporting teachers' peer learning of instructional gestures by teaching a mentee ai-agent together}.
\newblock \bibinfo{journal}{arXiv preprint arXiv:2505.17557} .
\bibitem[{Jin et~al.(2025)Jin, Yan, Echeverria, Ga{\v{s}}evi{\'c} and Martinez-Maldonado}]{jin2025generative}
\bibinfo{author}{Jin, Y.}, \bibinfo{author}{Yan, L.}, \bibinfo{author}{Echeverria, V.}, \bibinfo{author}{Ga{\v{s}}evi{\'c}, D.}, \bibinfo{author}{Martinez-Maldonado, R.}, \bibinfo{year}{2025}.
\newblock \bibinfo{title}{Generative ai in higher education: A global perspective of institutional adoption policies and guidelines}.
\newblock \bibinfo{journal}{Computers and Education: Artificial Intelligence} \bibinfo{volume}{8}, \bibinfo{pages}{100348}.
\bibitem[{Johnson and Johnson(1991)}]{johnson1991joining}
\bibinfo{author}{Johnson, D.W.}, \bibinfo{author}{Johnson, F.P.}, \bibinfo{year}{1991}.
\newblock \bibinfo{title}{Joining together: Group theory and group skills}.
\newblock \bibinfo{publisher}{Prentice-Hall, Inc}.
\bibitem[{Joo and Ko(2025)}]{joo2025ai}
\bibinfo{author}{Joo, S.H.}, \bibinfo{author}{Ko, E.G.}, \bibinfo{year}{2025}.
\newblock \bibinfo{title}{“[ai peers] are people learning from the same standpoint”: Perception of ai characters in a collaborative science investigation}, in: \bibinfo{booktitle}{International Conference on Artificial Intelligence in Education}, \bibinfo{organization}{Springer}. pp. \bibinfo{pages}{424--437}.
\bibitem[{Kamalov et~al.(2025)Kamalov, Calonge, Smail, Azizov, Thadani, Kwong and Atif}]{kamalov2025evolution}
\bibinfo{author}{Kamalov, F.}, \bibinfo{author}{Calonge, D.S.}, \bibinfo{author}{Smail, L.}, \bibinfo{author}{Azizov, D.}, \bibinfo{author}{Thadani, D.R.}, \bibinfo{author}{Kwong, T.}, \bibinfo{author}{Atif, A.}, \bibinfo{year}{2025}.
\newblock \bibinfo{title}{Evolution of ai in education: Agentic workflows}.
\newblock \bibinfo{journal}{arXiv preprint arXiv:2504.20082} .
\bibitem[{Khosravi et~al.(2022)Khosravi, Shum, Chen, Conati, Tsai, Kay, Knight, Martinez-Maldonado, Sadiq and Ga{\v{s}}evi{\'c}}]{khosravi2022explainable}
\bibinfo{author}{Khosravi, H.}, \bibinfo{author}{Shum, S.B.}, \bibinfo{author}{Chen, G.}, \bibinfo{author}{Conati, C.}, \bibinfo{author}{Tsai, Y.S.}, \bibinfo{author}{Kay, J.}, \bibinfo{author}{Knight, S.}, \bibinfo{author}{Martinez-Maldonado, R.}, \bibinfo{author}{Sadiq, S.}, \bibinfo{author}{Ga{\v{s}}evi{\'c}, D.}, \bibinfo{year}{2022}.
\newblock \bibinfo{title}{Explainable artificial intelligence in education}.
\newblock \bibinfo{journal}{Computers and education: artificial intelligence} \bibinfo{volume}{3}, \bibinfo{pages}{100074}.
\bibitem[{Kirsh(2009)}]{kirsh2009problem}
\bibinfo{author}{Kirsh, D.}, \bibinfo{year}{2009}.
\newblock \bibinfo{title}{Problem solving and situated cognition}, in: \bibinfo{booktitle}{The Cambridge handbook of situated cognition}. \bibinfo{publisher}{Cambridge University Press}, pp. \bibinfo{pages}{264--306}.
\bibitem[{Kulik and Fletcher(2016)}]{kulik2016effectiveness}
\bibinfo{author}{Kulik, J.A.}, \bibinfo{author}{Fletcher, J.D.}, \bibinfo{year}{2016}.
\newblock \bibinfo{title}{Effectiveness of intelligent tutoring systems: a meta-analytic review}.
\newblock \bibinfo{journal}{Review of educational research} \bibinfo{volume}{86}, \bibinfo{pages}{42--78}.
\bibitem[{Laal et~al.(2012)Laal, Laal and Kermanshahi}]{laal201221st}
\bibinfo{author}{Laal, M.}, \bibinfo{author}{Laal, M.}, \bibinfo{author}{Kermanshahi, Z.K.}, \bibinfo{year}{2012}.
\newblock \bibinfo{title}{21st century learning; learning in collaboration}.
\newblock \bibinfo{journal}{Procedia-Social and Behavioral Sciences} \bibinfo{volume}{47}, \bibinfo{pages}{1696--1701}.
\bibitem[{Lee et~al.(2025)Lee, Mun, Shin and Zhai}]{lee2025collaborative}
\bibinfo{author}{Lee, G.G.}, \bibinfo{author}{Mun, S.}, \bibinfo{author}{Shin, M.K.}, \bibinfo{author}{Zhai, X.}, \bibinfo{year}{2025}.
\newblock \bibinfo{title}{Collaborative learning with artificial intelligence speakers: pre-service elementary science teachers’ responses to the prototype}.
\newblock \bibinfo{journal}{Science \& Education} \bibinfo{volume}{34}, \bibinfo{pages}{847--875}.
\bibitem[{Lehtinen(2003)}]{lehtinen2003computer}
\bibinfo{author}{Lehtinen, E.}, \bibinfo{year}{2003}.
\newblock \bibinfo{title}{Computer-supported collaborative learning: An approach to powerful learning environments}.
\newblock \bibinfo{journal}{Powerful learning environments: Unravelling basic components and dimensions} \bibinfo{volume}{35}, \bibinfo{pages}{54}.
\bibitem[{Long and Magerko(2020)}]{long2020ai}
\bibinfo{author}{Long, D.}, \bibinfo{author}{Magerko, B.}, \bibinfo{year}{2020}.
\newblock \bibinfo{title}{What is ai literacy? competencies and design considerations}, in: \bibinfo{booktitle}{Proceedings of the 2020 CHI conference on human factors in computing systems}, pp. \bibinfo{pages}{1--16}.
\bibitem[{Ng et~al.(2021)Ng, Leung, Chu and Qiao}]{ng2021conceptualizing}
\bibinfo{author}{Ng, D.T.K.}, \bibinfo{author}{Leung, J.K.L.}, \bibinfo{author}{Chu, S.K.W.}, \bibinfo{author}{Qiao, M.S.}, \bibinfo{year}{2021}.
\newblock \bibinfo{title}{Conceptualizing ai literacy: An exploratory review}.
\newblock \bibinfo{journal}{Computers and Education: Artificial Intelligence} \bibinfo{volume}{2}, \bibinfo{pages}{100041}.
\bibitem[{Nguyen et~al.(2023)Nguyen, Ngo, Hong, Dang and Nguyen}]{nguyen2023ethical}
\bibinfo{author}{Nguyen, A.}, \bibinfo{author}{Ngo, H.N.}, \bibinfo{author}{Hong, Y.}, \bibinfo{author}{Dang, B.}, \bibinfo{author}{Nguyen, B.P.T.}, \bibinfo{year}{2023}.
\newblock \bibinfo{title}{Ethical principles for artificial intelligence in education}.
\newblock \bibinfo{journal}{Education and information technologies} \bibinfo{volume}{28}, \bibinfo{pages}{4221--4241}.
\bibitem[{O'Malley(2012)}]{o2012computer}
\bibinfo{author}{O'Malley, C.}, \bibinfo{year}{2012}.
\newblock \bibinfo{title}{Computer supported collaborative learning}. volume \bibinfo{volume}{128}.
\newblock \bibinfo{publisher}{Springer Science \& Business Media}.
\bibitem[{Ouyang and Jiao(2021)}]{ouyang2021artificial}
\bibinfo{author}{Ouyang, F.}, \bibinfo{author}{Jiao, P.}, \bibinfo{year}{2021}.
\newblock \bibinfo{title}{Artificial intelligence in education: The three paradigms}.
\newblock \bibinfo{journal}{Computers and Education: Artificial Intelligence} \bibinfo{volume}{2}, \bibinfo{pages}{100020}.
\bibitem[{Park et~al.(2023)Park, O'Brien, Cai, Morris, Liang and Bernstein}]{park2023generative}
\bibinfo{author}{Park, J.S.}, \bibinfo{author}{O'Brien, J.}, \bibinfo{author}{Cai, C.J.}, \bibinfo{author}{Morris, M.R.}, \bibinfo{author}{Liang, P.}, \bibinfo{author}{Bernstein, M.S.}, \bibinfo{year}{2023}.
\newblock \bibinfo{title}{Generative agents: Interactive simulacra of human behavior}, in: \bibinfo{booktitle}{Proceedings of the 36th annual acm symposium on user interface software and technology}, pp. \bibinfo{pages}{1--22}.
\bibitem[{Premack and Woodruff(1978)}]{premack1978does}
\bibinfo{author}{Premack, D.}, \bibinfo{author}{Woodruff, G.}, \bibinfo{year}{1978}.
\newblock \bibinfo{title}{Does the chimpanzee have a theory of mind?}
\newblock \bibinfo{journal}{Behavioral and brain sciences} \bibinfo{volume}{1}, \bibinfo{pages}{515--526}.
\bibitem[{Pu et~al.(2025)Pu, Lazaro, Arawjo, Xia, Xiao, Grossman and Chen}]{pu2025assistance}
\bibinfo{author}{Pu, K.}, \bibinfo{author}{Lazaro, D.}, \bibinfo{author}{Arawjo, I.}, \bibinfo{author}{Xia, H.}, \bibinfo{author}{Xiao, Z.}, \bibinfo{author}{Grossman, T.}, \bibinfo{author}{Chen, Y.}, \bibinfo{year}{2025}.
\newblock \bibinfo{title}{Assistance or disruption? exploring and evaluating the design and trade-offs of proactive ai programming support}, in: \bibinfo{booktitle}{Proceedings of the 2025 CHI Conference on Human Factors in Computing Systems}, pp. \bibinfo{pages}{1--21}.
\bibitem[{Reigeluth et~al.(2017)Reigeluth, Beatty and Myers}]{Reigeluth2017}
\bibinfo{editor}{Reigeluth, C.M.}, \bibinfo{editor}{Beatty, B.J.}, \bibinfo{editor}{Myers, R.D.} (Eds.), \bibinfo{year}{2017}.
\newblock \bibinfo{title}{Instructional-design theories and models, volume IV: The learner-centered paradigm of education}.
\newblock \bibinfo{publisher}{Routledge}.
\bibitem[{Roschelle and Teasley(1995)}]{roschelle1995construction}
\bibinfo{author}{Roschelle, J.}, \bibinfo{author}{Teasley, S.D.}, \bibinfo{year}{1995}.
\newblock \bibinfo{title}{The construction of shared knowledge in collaborative problem solving}, in: \bibinfo{booktitle}{Computer supported collaborative learning}, \bibinfo{organization}{Springer}. pp. \bibinfo{pages}{69--97}.
\bibitem[{Sapkota et~al.(2025)Sapkota, Roumeliotis and Karkee}]{sapkota2025ai}
\bibinfo{author}{Sapkota, R.}, \bibinfo{author}{Roumeliotis, K.I.}, \bibinfo{author}{Karkee, M.}, \bibinfo{year}{2025}.
\newblock \bibinfo{title}{Ai agents vs. agentic ai: A conceptual taxonomy, applications and challenges}.
\newblock \bibinfo{journal}{arXiv preprint arXiv:2505.10468} .
\bibitem[{Searle(1980)}]{searle1980minds}
\bibinfo{author}{Searle, J.R.}, \bibinfo{year}{1980}.
\newblock \bibinfo{title}{Minds, brains, and programs}.
\newblock \bibinfo{journal}{Behavioral and brain sciences} \bibinfo{volume}{3}, \bibinfo{pages}{417--424}.
\bibitem[{Shneiderman(2020)}]{shneiderman2020human}
\bibinfo{author}{Shneiderman, B.}, \bibinfo{year}{2020}.
\newblock \bibinfo{title}{Human-centered artificial intelligence: Reliable, safe \& trustworthy}.
\newblock \bibinfo{journal}{International Journal of Human--Computer Interaction} \bibinfo{volume}{36}, \bibinfo{pages}{495--504}.
\bibitem[{Strachan et~al.(2024)Strachan, Albergo, Borghini, Pansardi, Scaliti, Gupta, Saxena, Rufo, Panzeri, Manzi et~al.}]{strachan2024testing}
\bibinfo{author}{Strachan, J.W.}, \bibinfo{author}{Albergo, D.}, \bibinfo{author}{Borghini, G.}, \bibinfo{author}{Pansardi, O.}, \bibinfo{author}{Scaliti, E.}, \bibinfo{author}{Gupta, S.}, \bibinfo{author}{Saxena, K.}, \bibinfo{author}{Rufo, A.}, \bibinfo{author}{Panzeri, S.}, \bibinfo{author}{Manzi, G.}, et~al., \bibinfo{year}{2024}.
\newblock \bibinfo{title}{Testing theory of mind in large language models and humans}.
\newblock \bibinfo{journal}{Nature Human Behaviour} \bibinfo{volume}{8}, \bibinfo{pages}{1285--1295}.
\bibitem[{Sweller(2010)}]{sweller2010element}
\bibinfo{author}{Sweller, J.}, \bibinfo{year}{2010}.
\newblock \bibinfo{title}{Element interactivity and intrinsic, extraneous, and germane cognitive load}.
\newblock \bibinfo{journal}{Educational psychology review} \bibinfo{volume}{22}, \bibinfo{pages}{123--138}.
\bibitem[{Tirri et~al.(1999)Tirri, Husu and Kansanen}]{tirri1999epistemological}
\bibinfo{author}{Tirri, K.}, \bibinfo{author}{Husu, J.}, \bibinfo{author}{Kansanen, P.}, \bibinfo{year}{1999}.
\newblock \bibinfo{title}{The epistemological stance between the knower and the known}.
\newblock \bibinfo{journal}{Teaching and Teacher Education} \bibinfo{volume}{15}, \bibinfo{pages}{911--922}.
\bibitem[{Tomasello(2019)}]{tomasello2019becoming}
\bibinfo{author}{Tomasello, M.}, \bibinfo{year}{2019}.
\newblock \bibinfo{title}{Becoming human: A theory of ontogeny}.
\newblock \bibinfo{publisher}{Harvard University Press}, \bibinfo{address}{Cambridge, MA}.
\bibitem[{Tomasello et~al.(2005)Tomasello, Carpenter, Call, Behne and Moll}]{tomasello2005understanding}
\bibinfo{author}{Tomasello, M.}, \bibinfo{author}{Carpenter, M.}, \bibinfo{author}{Call, J.}, \bibinfo{author}{Behne, T.}, \bibinfo{author}{Moll, H.}, \bibinfo{year}{2005}.
\newblock \bibinfo{title}{Understanding and sharing intentions: The origins of cultural cognition}.
\newblock \bibinfo{journal}{Behavioral and brain sciences} \bibinfo{volume}{28}, \bibinfo{pages}{675--691}.
\bibitem[{Turkle(2011)}]{Turkle2011}
\bibinfo{author}{Turkle, S.}, \bibinfo{year}{2011}.
\newblock \bibinfo{title}{Alone together: Why we expect more from technology and less from each other}.
\newblock \bibinfo{publisher}{Basic Books}.
\bibitem[{Vaccaro et~al.(2024)Vaccaro, Almaatouq and Malone}]{vaccaro2024combinations}
\bibinfo{author}{Vaccaro, M.}, \bibinfo{author}{Almaatouq, A.}, \bibinfo{author}{Malone, T.}, \bibinfo{year}{2024}.
\newblock \bibinfo{title}{When combinations of humans and ai are useful: A systematic review and meta-analysis}.
\newblock \bibinfo{journal}{Nature Human Behaviour} \bibinfo{volume}{8}, \bibinfo{pages}{2293--2303}.
\bibitem[{Vygotsky(1978)}]{Vygotsky1978}
\bibinfo{author}{Vygotsky, L.S.}, \bibinfo{year}{1978}.
\newblock \bibinfo{title}{Mind in Society: The Development of Higher Psychological Processes}.
\newblock \bibinfo{publisher}{Harvard University Press}, \bibinfo{address}{Cambridge, MA}.
\bibitem[{Wang et~al.(2025)Wang, Wang, Chen, Liu, Bao and Xu}]{wang2025impact}
\bibinfo{author}{Wang, H.}, \bibinfo{author}{Wang, C.}, \bibinfo{author}{Chen, Z.}, \bibinfo{author}{Liu, F.}, \bibinfo{author}{Bao, C.}, \bibinfo{author}{Xu, X.}, \bibinfo{year}{2025}.
\newblock \bibinfo{title}{Impact of ai-agent-supported collaborative learning on the learning outcomes of university programming courses}.
\newblock \bibinfo{journal}{Education and Information Technologies} , \bibinfo{pages}{1--33}.
\bibitem[{Wang et~al.(2024)Wang, Ma, Feng, Zhang, Yang, Zhang, Chen, Tang, Chen, Lin et~al.}]{wang2024survey}
\bibinfo{author}{Wang, L.}, \bibinfo{author}{Ma, C.}, \bibinfo{author}{Feng, X.}, \bibinfo{author}{Zhang, Z.}, \bibinfo{author}{Yang, H.}, \bibinfo{author}{Zhang, J.}, \bibinfo{author}{Chen, Z.}, \bibinfo{author}{Tang, J.}, \bibinfo{author}{Chen, X.}, \bibinfo{author}{Lin, Y.}, et~al., \bibinfo{year}{2024}.
\newblock \bibinfo{title}{A survey on large language model based autonomous agents}.
\newblock \bibinfo{journal}{Frontiers of Computer Science} \bibinfo{volume}{18}, \bibinfo{pages}{186345}.
\bibitem[{Ward et~al.(2011)Ward, Nolen and Horn}]{ward2011productive}
\bibinfo{author}{Ward, C.J.}, \bibinfo{author}{Nolen, S.B.}, \bibinfo{author}{Horn, I.S.}, \bibinfo{year}{2011}.
\newblock \bibinfo{title}{Productive friction: How conflict in student teaching creates opportunities for learning at the boundary}.
\newblock \bibinfo{journal}{International Journal of Educational Research} \bibinfo{volume}{50}, \bibinfo{pages}{14--20}.
\bibitem[{Wei et~al.(2025)Wei, Wang, Lee and Liu}]{wei2025effects}
\bibinfo{author}{Wei, X.}, \bibinfo{author}{Wang, L.}, \bibinfo{author}{Lee, L.K.}, \bibinfo{author}{Liu, R.}, \bibinfo{year}{2025}.
\newblock \bibinfo{title}{The effects of generative ai on collaborative problem-solving and team creativity performance in digital story creation: an experimental study}.
\newblock \bibinfo{journal}{International Journal of Educational Technology in Higher Education} \bibinfo{volume}{22}, \bibinfo{pages}{23}.
\bibitem[{Weijers et~al.(2025)Weijers, Wu, Betts, Jacod, Guan, Sujaya, Dev, Goel, Delooze, Rabbany et~al.}]{weijers2025intuition}
\bibinfo{author}{Weijers, R.}, \bibinfo{author}{Wu, D.}, \bibinfo{author}{Betts, H.}, \bibinfo{author}{Jacod, T.}, \bibinfo{author}{Guan, Y.}, \bibinfo{author}{Sujaya, V.}, \bibinfo{author}{Dev, K.}, \bibinfo{author}{Goel, T.}, \bibinfo{author}{Delooze, W.}, \bibinfo{author}{Rabbany, R.}, et~al., \bibinfo{year}{2025}.
\newblock \bibinfo{title}{From intuition to understanding: Using ai peers to overcome physics misconceptions}.
\newblock \bibinfo{journal}{arXiv preprint arXiv:2504.00408} .
\bibitem[{Wellman et~al.(1990)Wellman, Carey, Gleitman, Newport and Spelke}]{wellman1990child}
\bibinfo{author}{Wellman, H.M.}, \bibinfo{author}{Carey, S.}, \bibinfo{author}{Gleitman, L.}, \bibinfo{author}{Newport, E.L.}, \bibinfo{author}{Spelke, E.S.}, \bibinfo{year}{1990}.
\newblock \bibinfo{title}{The child's theory of mind}.
\newblock \bibinfo{publisher}{The MIT Press}.
\bibitem[{Xi et~al.(2025)Xi, Chen, Guo, He, Ding, Hong, Zhang, Wang, Jin, Zhou et~al.}]{xi2025rise}
\bibinfo{author}{Xi, Z.}, \bibinfo{author}{Chen, W.}, \bibinfo{author}{Guo, X.}, \bibinfo{author}{He, W.}, \bibinfo{author}{Ding, Y.}, \bibinfo{author}{Hong, B.}, \bibinfo{author}{Zhang, M.}, \bibinfo{author}{Wang, J.}, \bibinfo{author}{Jin, S.}, \bibinfo{author}{Zhou, E.}, et~al., \bibinfo{year}{2025}.
\newblock \bibinfo{title}{The rise and potential of large language model based agents: A survey}.
\newblock \bibinfo{journal}{Science China Information Sciences} \bibinfo{volume}{68}, \bibinfo{pages}{121101}.
\bibitem[{Yan et~al.(2025a)Yan, Greiff, Lodge and Ga{\v{s}}evi{\'c}}]{yan2025distinguishing}
\bibinfo{author}{Yan, L.}, \bibinfo{author}{Greiff, S.}, \bibinfo{author}{Lodge, J.M.}, \bibinfo{author}{Ga{\v{s}}evi{\'c}, D.}, \bibinfo{year}{2025}a.
\newblock \bibinfo{title}{Distinguishing performance gains from learning when using generative ai}.
\newblock \bibinfo{journal}{Nature Reviews Psychology} , \bibinfo{pages}{1--2}.
\bibitem[{Yan et~al.(2024)Yan, Greiff, Teuber and Ga{\v{s}}evi{\'c}}]{yan2024promises}
\bibinfo{author}{Yan, L.}, \bibinfo{author}{Greiff, S.}, \bibinfo{author}{Teuber, Z.}, \bibinfo{author}{Ga{\v{s}}evi{\'c}, D.}, \bibinfo{year}{2024}.
\newblock \bibinfo{title}{Promises and challenges of generative artificial intelligence for human learning}.
\newblock \bibinfo{journal}{Nature Human Behaviour} \bibinfo{volume}{8}, \bibinfo{pages}{1839--1850}.
\bibitem[{Yan et~al.(2025b)Yan, Martinez-Maldonado, Jin, Echeverria, Milesi, Fan, Zhao, Alfredo, Li and Ga{\v{s}}evi{\'c}}]{yan2025effects}
\bibinfo{author}{Yan, L.}, \bibinfo{author}{Martinez-Maldonado, R.}, \bibinfo{author}{Jin, Y.}, \bibinfo{author}{Echeverria, V.}, \bibinfo{author}{Milesi, M.}, \bibinfo{author}{Fan, J.}, \bibinfo{author}{Zhao, L.}, \bibinfo{author}{Alfredo, R.}, \bibinfo{author}{Li, X.}, \bibinfo{author}{Ga{\v{s}}evi{\'c}, D.}, \bibinfo{year}{2025}b.
\newblock \bibinfo{title}{The effects of generative ai agents and scaffolding on enhancing students’ comprehension of visual learning analytics}.
\newblock \bibinfo{journal}{Computers \& Education} , \bibinfo{pages}{105322}.
\bibitem[{Yan et~al.(2025c)Yan, Pammer-Schindler, Mills, Nguyen and Ga{\v{s}}evi{\'c}}]{yan2025beyond}
\bibinfo{author}{Yan, L.}, \bibinfo{author}{Pammer-Schindler, V.}, \bibinfo{author}{Mills, C.}, \bibinfo{author}{Nguyen, A.}, \bibinfo{author}{Ga{\v{s}}evi{\'c}, D.}, \bibinfo{year}{2025}c.
\newblock \bibinfo{title}{Beyond efficiency: Empirical insights on generative ai's impact on cognition, metacognition and epistemic agency in learning}.
\bibitem[{Y{\i}ld{\i}z(2025)}]{yildiz2025minds}
\bibinfo{author}{Y{\i}ld{\i}z, T.}, \bibinfo{year}{2025}.
\newblock \bibinfo{title}{The minds we make: A philosophical inquiry into theory of mind and artificial intelligence}.
\newblock \bibinfo{journal}{Integrative Psychological and Behavioral Science} \bibinfo{volume}{59}, \bibinfo{pages}{10}.
\bibitem[{Yusuf et~al.(2025)Yusuf, Money and Daylamani-Zad}]{yusuf2025pedagogical}
\bibinfo{author}{Yusuf, H.}, \bibinfo{author}{Money, A.}, \bibinfo{author}{Daylamani-Zad, D.}, \bibinfo{year}{2025}.
\newblock \bibinfo{title}{Pedagogical ai conversational agents in higher education: a conceptual framework and survey of the state of the art}.
\newblock \bibinfo{journal}{Educational technology research and development} \bibinfo{volume}{73}, \bibinfo{pages}{815--874}.
\bibitem[{Zhang et~al.(2021)Zhang, McNeese, Freeman and Musick}]{zhang2021ideal}
\bibinfo{author}{Zhang, R.}, \bibinfo{author}{McNeese, N.J.}, \bibinfo{author}{Freeman, G.}, \bibinfo{author}{Musick, G.}, \bibinfo{year}{2021}.
\newblock \bibinfo{title}{" an ideal human" expectations of ai teammates in human-ai teaming}.
\newblock \bibinfo{journal}{Proceedings of the ACM on Human-Computer Interaction} \bibinfo{volume}{4}, \bibinfo{pages}{1--25}.
\bibitem[{Zhou and Schofield(2024)}]{zhou2024using}
\bibinfo{author}{Zhou, X.}, \bibinfo{author}{Schofield, L.}, \bibinfo{year}{2024}.
\newblock \bibinfo{title}{Using social learning theories to explore the role of generative artificial intelligence (ai) in collaborative learning}.
\newblock \bibinfo{journal}{Journal of Learning Development in Higher Education} .

\end{thebibliography}


\appendix
\end{document}